\input epsf.tex
\documentstyle[12pt]{article}

\newcommand{\be}{\begin{equation}}   \newcommand{\ee}{\end{equation}}
\newcommand{\bear}{\begin{eqnarray}}
\newcommand{\eear}{\end{eqnarray}}
\newcommand{\ba}{\begin{array}}      \newcommand{\ea}{\end{array}}

\begin{document}
\pagestyle{empty}
\begin{titlepage}

\vspace*{-8mm}
\noindent 
\makebox[11.5cm][l]{BUHEP-96-31} October 17, 1996\\
\makebox[11.5cm][l]{} Revised January 30, 1997 \\
\vspace{2.cm}
\begin{center}
  {\LARGE {\bf  Estimating Vacuum Tunneling Rates  }}\\
\vspace{42pt}
Indranil Dasgupta \footnote{e-mail address:
dgupta@budoe.bu.edu}

\vspace*{0.5cm}

 \ \ Department of Physics, Boston University \\
{590 Commonwealth Avenue, Boston, MA 02215, USA}

\vskip 3.4cm
\end{center}
\baselineskip=18pt

\begin{abstract}
{\normalsize
We show that in Euclidean field theories that have bounce solutions,
the bounce with the least action is 
the {\it {global}} minimum of the action in
an open space of field configurations. 
A rigorous upper bound on the minimal bounce action can 
therefore be obtained by finite numerical methods. This sets a lower
bound on the tunneling rate which, fortunately, is often the more
interesting and useful 
bound. We introduce a notion of reduction which allows
this bound to be computed with less effort by reducing
complicated field theories to simpler ones.
}

\end{abstract}

\vfill
\end{titlepage}

\baselineskip=18pt  
\pagestyle{plain}
\setcounter{page}{1}


\section {Introduction}

Vacuum tunneling in field theories was treated semiclassically in ref. 
\cite {coleman}. In the small $\hbar$ limit the tunneling rate per unit
volume can be
expanded as:
\be
{\Gamma \over V}= A\, {\rm {exp}}\left 
( {-S[ \overline{\phi} ]\over \hbar} \right )
 \times \left [1 +O(\hbar) \right ] \; .
\label {transition}
\ee
The exponent is the Euclidean action of the saddle point configuration
called the ``bounce''. The symbol $\overline {\phi}$ denotes the entire
field configuration at the bounce. The prefactor $A$ has the dimensions of 
${\rm {(mass)}}^d$ in a $d$ dimensional theory 
and can be formally written as 
a ratio of two operator determinants. 
The tunneling rate is usually 
estimated by replacing $A$ by $m^d$, where
$m$ is a characteristic mass scale of the theory. The error in making 
this approximation is not significant compared to the error coming 
from the exponential term which is determined by numerically computing the
bounce action $S[ \overline{\phi} ]$ (a more accurate estimate may be
obtained in some cases, see ref. \cite {wipf1}).

The determination of ${\Gamma / V}$ is important in extensions of
the standard model where the effective potential may have several
inequivalent vacua and the realistic and viable vacuum may turn out to
be only a local minimum of the potential. In that case, we would be
living in a false vacuum now, and our very existence would imply some
upper bound on the tunneling rate. 
If a lower bound on the tunneling
rate can be obtained by analytical or numerical methods, then a
comparison with the {\it {experimental}} upper bound 
may place strong constraints on the parameters of the
theory. One is therefore often interested in finding an upper bound on
the bounce action by analytical or numerical methods.

Recently there has been a renewed interest
in considering the tunneling efects in models of particle physics \cite
{ddr, klw, ccb, kls, ak}. 
The effect of such tunneling was considered 
in ref. \cite {ddr} to constrain the parameter space of the
existing models of dynamical supersymmetry breaking (DSB). 
Similar tunneling rate computations exist in the
literature \cite {ccb,kls,ak} 
in the context of the minimal supersymmetric standard model
(MSSM), where 
charge and color breaking (CCB) vacuua exist \cite {stab}. 
While analytical estimates of the bounce action can be made in special cases
\cite {coleman,wipf2}, the realistic models of DSB and MSSM often
fall in the intractable cases where one must resort to 
numerical methods. When several fields and coupling constants are
involved, the numerical search for the bounce 
becomes long and time consuming. 
The search also needs to be repeated many times to map a significant region
of the parameter space. New techniques for approximating the bounce
action in a short time are therefore welcome.

In this paper we 
point out the limitations of existing techniques of approximation
and indicate a direction
for further simplifications and improvements. In particular 
we show that all finite numerical methods are useful only to find an
upper bound on the bounce action. We then develop a notion of
``reduction'' that provides an intuitive and systematic approach to
better techniques for obtaining this bound.

\section {The Bounce as a Local Minimum}

To begin with, let us consider a theory with $m$ scalar fields in $d$
dimensions. The Euclidean action is
\bear
S = T + V ;
\label {stv}
\eear
where $T= \sum _{a=1}^m
{1 \over 2} \int d^dx\left ({\partial \phi_a \over \partial
  x_i}\right )^2 $ and 
$V= \int d^dx\, U\left(\phi_1, \phi_2 ..\phi_m\right )$
with $i=1,2...d$ and $a=1,2...m$. Suppose the potential $ U(\phi_1,
\phi_2 ...\phi_m)$ has a local minimum \footnote {By a
suitable redefinition of the fields, the local minimum can be made to
lie at the point $\phi_a =0$.} at $\phi_a = 0$ and a 
global minimum at $\phi_a = \phi_a^t$. Then a 
bounce solution $\overline {\phi}_a(x)$ must 
possess the following three properties:
(i) it approaches the false vacuum $\overline {\phi_a} = 0$ 
at $t \to \pm \infty$;
(ii) it is a saddle point of the action with a single ``direction of
instability'';
(iii) at the turning point (say at $t=0$), all the generalized
velocities $ {\partial \overline {\phi_a} \over \partial x_i}$ are zero.

It will be useful to write down the effect of the 
scale transformation ($x \rightarrow \lambda x$) on  $T$ and $V$. In $d$
dimensions one has: 
$T \rightarrow T_{\lambda}=\lambda ^{d-2}T$ and 
$ V \rightarrow V_{\lambda}=\lambda ^dV$.
Since the bounce is an extremum of the action, we
must have, 
${d \over d \lambda} S_{\lambda}(\overline \phi)|_{\lambda = 1}= 0$, 
from which it follows that
\be \label {tv}
(d-2)T({\overline \phi}) + (d)V({\overline \phi}) = 0 \; .
\ee
{From} (\ref {tv}) 
one also finds that ${d^2 \over d \lambda ^2} S_{\lambda}(\overline
\phi)|_{\lambda = 1 } = (d-2)(d-3)T({\overline \phi})
+d(d-1)V({\overline \phi}) <0$ when $d>2$. 
Therefore the direction of instability associated with the bounce has a
component 
corresponding  to the scale transformations for $d>2$. 

In \cite {cmg} it was shown that in $d> 2$ dimensional 
theories with a {\it {single real scalar field}} ($a=1$), the bounce 
solution with the least action 
is an $O(d)$ invariant field configuration with the 
boundary conditions:
${d\overline {\phi_1}(r) \over dr}|_{r=0}=0$ and 
$\overline {\phi_1}(r)|_{r \to \infty}=0 $, 
where $r$ is the radial coordinate in $R^d$.
In this case, the task of
finding the bounce action numerically is rather easy.
One simply solves the Euclidian equations of
motion from some initial point $\phi_1 (0)$ and looks at the limiting
value
$\phi_1 (\infty)$. For arbitrary values of $\phi_1 (0)$ the value of $\phi_1
(\infty)$ is either an ``overshoot'' ($\phi_1 (\infty)>0$) or an
``undershoot'' ($\phi_1 (\infty)<0$). Since the correct value of
$\phi_1 (0)
$ (called the ``escape point'')
must lie between two trial values which end in an overshoot and an
undershoot, the search converges rapidly by bisections. 

The task is considerably 
more complicated in the case of theories with 
many scalar fields. The bracketing property of the overshoot 
and the undershoot, which is
obviously valid only when the field is a real scalar quantity, is
lost. Therefore one can not search for the escape point by integrating
the equations of motion.
Because the bounce is
not a minimum of the action, but a saddle point, a simple minimization
of the action would not work either.
For instance, the method suggested in 
ref. \cite {cl} of 
maximizing  the action of a field configuration 
on the lattice with respect to scale transormations and then minimizing
it with repect to random variations of the fields
is unlikely to converge to a bounce solution because 
variations that are orthogonal to a scale transformation are hard to
identify in an actual lattice computation. A remarkable simplification
is however achieved by adopting the method of 
ref. \cite {kls}
which consists of adding to the Euclidean
action other terms that have the following two properties: \\
(i) they vanish at the bounce,\\
(ii) they remove the instability associated with scaling, i.e, the
``improved'' action with the new terms is minimized with respect to {\it
  {all}} variations including the scale transformation. 

When terms like these are added to the action, a bounce appears as a
local minimum (instead of a saddle point) of the improved action.
An example of the improved action is 
\be \label {improved}
\overline {S}(\phi_a)= {S(\phi_a)}+ \Sigma _n \alpha _n |\Lambda|^{p_n} \; ,
\ee
where $\Lambda = (d-2)T + (d)V$ 
and $p_n$ and $\alpha _n$ are positive numbers for
$n = 1, 2 ... \;$. For $d>2$ 
the new terms added to the action have the two properties
mentioned above \cite {kusenko}.

Although the method of improved action 
reduces the problem to a pure minimization, the
bounce is only a local minimum in an 
infinite dimensional space. In the absence of further information,
it is by no means obvious that minimizations by finite numerical methods
actually yield a useful result. 
Typically, the discretization involved in the 
numerical methods generates spurious minima
of the improved action. It is not clear if the result of such
a search yields an action that is greater or less than the bounce action.
This is a drawback
since one is often interested in 
a rigorous {\it {upper bound}} on the bounce action. 
Although finite numerical methods will always have some limitations
(which we discuss later) we will now 
show that they can be sufficient to obtain such a rigorous upper bound. The
approach suggested by us will also lead to new avenues of
simplification and refinement of numerical techniques.

\section {The Spherical Bounce as a Global Minimum}

Let us define a $spherical$ bounce as an $O(d)$ symmetric
field configuration with
properties (i) and (ii) of the bounce provided the variations are
restricted to preserve the $O(d)$ invariance. Clearly the $O(d)$ invariance of
the configuration implies that it has property (iii) of the bounce.
Our approach rests on the following two observations about the spherical
bounces. \footnote {This approach is closely related to the one suggested
by A. Wipf \cite {wipf2}.}

{\bf Statement 1}: A spherical bounce is also a true bounce.

{\bf Statement 2}: The spherical bounce with the least action is a
{\it {global}} minimum of the action in an open space of field configurations
which obey the boundary conditions appropriate for a bounce.

The proof of Statement 2 is simpler
 for $d>2$. We will discuss the cases $d=1,2$ later. 
Before we sketch the proofs, let us first point out some immediate gains
that result from the knowledge of the space $C_1$. At the heart of our
aproach is the trivial point that it is always easier to bound a {\it
{global}} minimum from above than a {\it {local}} minimum. 

There are two 
general limitations associated with finite numerical
methods. Firstly, in multi-scalar theories there may be several bounce
solutions. The tunneling rate is determined by the bounce (or bounces)
with the least action. There is no numerical method that guarantees
convergence to the bounce with the least action. Secondly, any method
of discretizing the improved action (\ref {improved}) introduces
spurious local minima which are local minima of the {\it {discretized}}
improved action with respect to a finite dimensional space of
variations, but are not local minima when all continuous variations of
the fields are considered. These spurious minima 
may convey no information about any bounce action. 

These 
limitations preclude the possibility of finding the least bounce action
to within
controllable numerical errors in multi-scalar theories. 
However, as we have emphasized earlier,
in many cases it is useful to find just an upper bound on the
least bounce action. The knowledge of the space $C_1$ allows one to
approximate the bounce by a field configuration that lies {\it
  {exactly}} on the space $C_1$ and provides a {\it {rigorous}} upper bound on
the least spherical bounce action (which itself bounds from above the
least bounce action). As it happens the space $C_1$ is quite
simple and it is easy to find field configurations lying on it. 
We now present the proofs of Statements 1 and 2. The latter also defines
the space $C_1$.

{\bf {Proof of Statement 1}}: This statement is in fact a consequence of
the principle of symmetric criticality \cite {palais}. 
Writing the action of (\ref {tv}) 
in polar coordinates we
  have 
\bear
S &=& S_1 + S_2 \; , \nonumber \\ 
S_1&=& \int d\Omega \, r^{d-1}dr \left [ {1\over 2} \sum  _{a=1}^m
\left ({d\phi_a \over dr}\right )^2
+ U(\phi_1,\phi_2 ... \phi_m) \right ] \; ,\nonumber \\
S_2&=& \int d\Omega \, r^{d-1}dr  \left [ {1\over 2r^2} \sum _{a=1}^m  
\sum  _{i=1}^{d-1}
f_i(\Omega)^2\left ({d\phi_a \over d \theta_i}\right )^2 \right ]\; , 
\label {polar}
\eear
where $\theta_i$ are the angular coordinates and $f_i$ are the
measures corresponding to the angular gradients ${d\phi_a \over d
  \theta_i}$. The quantity $S_2$ is
positive definite and is minimized on the space of $O(d)$ invariant
configurations. Therefore a spherical bounce is an extremum with respect to  the $O(d)$
breaking variations too and no new negative eigenvalue is added to the
Hessian $\partial^2 S \over \partial \phi_a^2$ by considering the $O(d)$
breaking variations. That is, the spherical bounce is also a true
bounce. {\bf {Q.E.D.}}

{\bf {Proof of Statement 2}} (for $d>2$): 
Consider the space of all $O(d)$ invariant field configurations that satisfy
the correct boundary conditions and are extremized with respect to  the scale
transformations $x\to \lambda x$. Let us call this space $C$. All points
in $C$ obey (\ref {tv}). The space $C$ naturally splits into
two disconnected parts $C=C_0 \oplus C_1$, where $C_0$ is the 
connected space containing the
trivial solution $\phi_a \equiv 0$ where all fields assume their values
at the 
the false vacuum at all times. 
At the trivial solution $V=0$, and in a small ball
around this solution $V>0$ because the false vacuum is a local minimum
of energy. On the other hand from (\ref {tv}) it is clear that 
for points in $C$, $V \le 0$. Thus $C_0$
consists of an isolated point.

Points in $C_1$ have $V<0$ and $T>0$. Every nontrivial
configuration satisfying (\ref {tv}) belongs to $C_1$. But the the point
$G$ which is the global
minimum of the action in $C_1$ satisfies the correct boundary conditions
 for the bounce, is a saddle point with a single direction of
 instability (the direction corresponding to the generator of scale
transformations which is ``orthogonal'' to $C_1$) and
has ${\partial \phi_a \over \partial t}=0$ at $t=0$ by the $O(d)$
invariance. Therefore, by definition, it is the spherical bounce with the 
least action. {\bf Q.E.D.}

{\bf Comment 1:} It is obvious that the action of an arbitrary field 
configuration in the space $C$ is necessarily bounded below by either
the action of the 
trivial solution $C_0$ or the action of a nontrivial solution of the
equations of motion lying on $C_1$. It is a remarkable property of the
space $C$ that it is always the latter.
The essential point is this. In a discrete numerical method one 
allows for variations of the action over a finite dimensional
space only. If scale transformations are included in the variations and 
$O(d)$ invariance is imposed throughout the search,
then one is assured that the ``bounce'' obtained by the numerical method
lies on the space $C_1$ and provides the required upper bound. 

{\bf {Comment 2:}} The $O(d)$ symmetry is
indispensible for any numerical method because it reduces the number of
variables drastically. However it is 
redundant from the point of
view of defining the space $C_1$. A milder symmetry like the $Z_2$
symmetry $\phi_a(t,{\bf {x}})= \phi_a(-t,{\bf {x}})$ is sufficient
to ensure that $G$ has property (iii) of the bounce. One can redefine
the spaces $C$ and $C_1$ with the $O(d)$ symmetry replaced by the 
$Z_2$ symmetry.{\it { The corresponding point $G$ is then truely the
bounce with the minimum action}}. This is because, as one can easily show,
the point $G$ is a true bounce and, conversely, every true bounce 
has the $Z_2$ symmetry possessed by $G$. The latter assertion follows
from observing that the equations of motion are invariant under $t \to -t$ and
imposing ${\partial \overline {\phi_a} 
\over \partial t }(t=0)\equiv 0$ implies that any solution $\overline
{\phi_a}$ of the equations of motion also has this symmetry.

{\bf Comment 3:} There is more to gain from the above approach 
 than the insight
that numerical methods are useful only to find upper bounds on the
bounce action. New and more effective techniques are suggested. 
This is motivated as follows.
Underlying any numerical technique is a discretization of the
action. For instance a straightforward discretization of the improved
action with $O(d)$ symmetry is:
\bear
\overline {S}(\phi) & = & T(\phi) + V(\phi) + \alpha |T(\phi) + 2
V(\phi)|^2 \nonumber \\
T(\phi) & = & \Omega \Delta ^d \sum _{n=1}^N \sum_{a=1}^m n^3
 {[\phi _a ^{(n+1)} -\phi _a ^{(n)}]^2 \over 2 \Delta ^2 } \nonumber \\
V(\phi) & = & \Omega \Delta ^d \sum _{n=1}^N n^3 U(\phi _1 ^{(n)},
\phi _2 ^{(n)},...\phi _m ^{(n)}) \; .
\label {lattice}
\eear
The radial coordinate in $R^d$ is discretized into $N$ points and
$\Omega $ is the solid angle in $d$ dimensions. The
purpose of the discretization is to 
{\it {reduce}} the space of field variations to a finite
dimensional space (to a space $C_f$ here, which is
homomorphic to $R^{N\times m}$). 

There is no particular reason to believe that the discretization
in (\ref {lattice}) is the most convenient way of reducing the space
of all variations
 to a finite dimensional space. Indeed, as we show below, tunneling in theories with many
scalar fields can be extremely difficult to explore with the
discretization of (\ref {lattice}). Since the most significant part of
the computations is to make the final point lie on the space $C_1$ and
not how the reduction to some space $C_f$ is achieved, 
the natural question to ask is: why not implement the ``reduction'' of
allowed variations  at an
early stage by reducing the complicated field theory to a simpler one? 
The question itself suggests new and systematic ways of improving the
numerical techniques.

\section {Methods of Reduction}

We will seek new methods of
reduction. Our search is guided by the technical difficulties
encountered with the 
discretization of (\ref {lattice}). Let us briefly describe 
the essential features and the limitations of the 
reduction characterizing this discretization. We will call this method
of reduction method 1.
        
{\bf {Method 1:}} With the discretization (\ref {lattice}) one minimizes
the action with the constraint $(d-2)T+(d)V=0$. The constraint
can be enforced by making the Lagrange multiplier $\alpha$ large.
The result is a discrete
trajectory $\overline {\phi}_a(n)$ with $n=1, 2 ... N$,
which can be extended to the
continuous and piecewise linear function $\overline {\phi}_a(r)$ given
by $\overline {\phi}_a(r)= \left [(1-r+n)
\overline {\phi}_a(n)+(r-n)\overline {\phi}_a(n+1)\right]$ for $n\le r < n+1$
and  $\overline {\phi}_a(r) =\overline {\phi}_a(N)$ for $r>N$. When $U$
is a polynomial in $\phi_a$, the 
functions $T$ and $V$ can be calculated to arbitrary accuracy using this
function. The condition (\ref {tv}) may not be exactly satisfied by $T$
and $V$ obtained at this point. But one can always 
perform a scale transformation to satisfy  (\ref {tv}) exactly at the
end. Thus the method of improved action is successful in bounding
the bounce action provided one performs the necessary scale
transformation at the end. 

In this case the search space for the minimization 
is $N \times m$ dimensional, with $m$
fixed by the theory. In an
$N \times m$ dimensional Euclidean space, if one desires to reach within a 
distance of
$\delta $ from a minimum by a random iterative
search (as suggested in ref.\cite {kls}), the number of steps
required is about $({l \over \delta})^{Nm}$, where $l$ is the maximum
step length in any direction. One can improve this by 
using a multidimensional 
 ``greedy'' 
minimization technique
such as the Conjugate Gradient Method \cite {recipes}. The computational
time typically grows as some power of $N \times m$ (depending on the
complexity of the function). Also, the search space has many local
minima with widely varying actions. 
Usually, one needs several iterations to arrive at a good one. Clearly
the computation becomes harder as the number of fields $m$ increases. 
Therefore a 
 method of reduction where the computational complexity and time do
not depend on $m$ becomes desirable for 
theories with many fields. The alternative 
method of reduction presented below has precisely this virtue.

{\bf {Method 2:}} The action is given by (\ref {stv}). 
The false vauum is at $\phi_a=0$. Choose a straight line, passing
through the point $\phi_a=0$, in the
$m$ dimensional Euclidean space $R^m$ with axes given by $\phi_1, \phi_2
..\phi_m$.
The fields $\phi_a$ can be constrained to take values 
only on this straight line by putting $\phi_a = y_a\phi$ where
$y_a$ are real numbers staisfying $\sum y_a^2=1$ and $\phi$ is the
{\it {reduced}} scalar field. Then the action is reduced to the {\it 
{reduced}} action $S[\phi]= \int d^dx \left 
[1/2({\partial \phi \over \partial
  x_{i}} )^2 + U (\phi)\right ]$ which is a functional of 
a single real scalar field and the bounce
action in this reduced theory 
is easily computed by the method of bisections described earlier.
Note that the bounce exists if and only if the chosen 
straight line passes through some point $\phi_a^l$ such that $U\left
(0 \right ) > U\left(\phi _a^l \right ) $. 
This method is an obvious 
reduction that intuitively seems to be a correct
simplification. However, the relation of the ``bounce'' obtained in the
reduced theory
to the true bounce may not be immediately clear. 
But note that the action is extremized 
with respect to  all variations that do not move the fields $\phi_a(x)$ out of the
chosen straight line in field space. The crucial point is that scale
transformations are included in these variations. Therefore the solution
to the equation of motion 
satisfies (\ref {tv}) and by 
Statement 2 its action 
is rigorously an upper bound
on the action of the bounce with least action in the full
theory. The search time scales like $N^{{\rm {log}}\left ({|\phi^t - \phi^f|
\over \delta}\right )}$ when the search is made by the method of undershoots and
overshoots, and has no dependence on $m$. The problem of multiple
bounces and spurious minima does not usually arise. Also, the result is
automatically a point on $C_1$ without the need for adding new terms to
the action as in (\ref {improved}).

In practice one encounters polynomial potentials in $\phi_a(x)$. It is
often possible to find the position of the true vacuum or the saddle
point in the potential between the true vacuum and the false vacuum. The
lines joining these points to the false vacuum may yield fairly low
values for the upper bound on the 
bounce action. One can do the search over several lines
which can be judiciously chosen.

\vspace{-0.5cm}
\centerline{\epsfxsize=2.0in\epsfbox{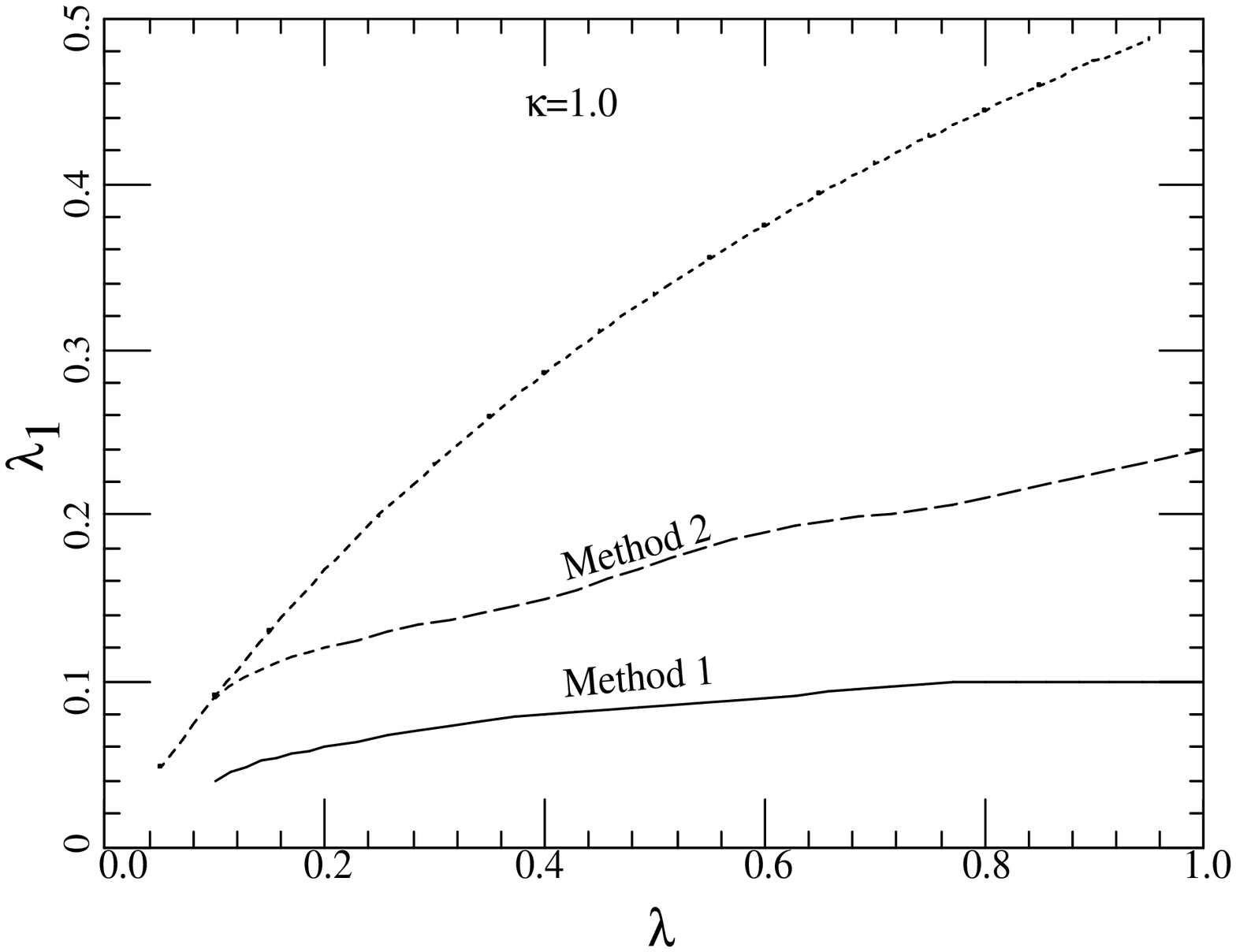}}
\vspace{-1.0cm}
\makebox[0.8in][l]{\hspace{2ex} Fig. 1.}
\parbox[t]{4.8in}{ {\small Comparison of Methods 1 and 2.
The region above the dashed (solid) line is ruled out 
by method 1 (method 1). } }
\bigskip

In Figure 1 we show a comparison of the performance of the above two
methods in a realistic 
DSB model \cite {dn} whose vacuum stability
has been studied in ref.\cite {ddr}. For our purpose it is sufficient to
consider the scalar fields in the so called messenger sector of the
theory which serves to communicate supersymmetry breaking to the fields
in the standard model. For a natural range of parameters, the
true vacuum in the messenger sector has been shown to be 
color breaking \cite {ddr}.
The relevant part of the action consists of the complex
fields $P, N, S, q$ and $ \overline {q}$. The last two fields carry color
but the color index can be suppressed in this discussion. 
Apart from the
usual kinetic energy terms, the action has the potential:
\bear
V&=& {g^2\over
  2}(|P|^2-|N|^2)+(M_2^2+\lambda_1^2|S|^2)(|P|^2+|N|^2)\\ \nonumber
&+& \kappa ^2|S|^2(|q|^2+|\overline {q}|^2)+|\kappa q\overline {q}+
\lambda S^2 + \lambda_1PN|^2 + \alpha ^2 (|P|^2+|N|^2)^2\; , 
\label {dsbv}
\eear
where $g, \lambda , \lambda_1$ and $\kappa$ are positive coupling
constants, $M_2^2<0$ and $\alpha^2$ is of the order of $10^{-2}$. 
A false but phenomenologically viable vacuum exists at $q=\overline {q}=0$
and $|P|=|N|$ with $S\ne 0$ if $\lambda_1 < (\kappa \lambda_1)/(\kappa
+\lambda)$ \cite {ddr}. We have computed
the vacuum tunneling rate using both the methods described above. In
Figure 1 we plot our results in the $\lambda , \lambda_1$ plane
for $\kappa = 1.0$ and $\alpha ^2=0.01$. 
The region below the dotted line is allowed analytically, but 
the region above the dashed (solid) curve is ruled out by method 2 (method
1) because the lifetime of the false vacuum is shorter than the age of
the universe. In this case
a larger region in the 
parameter space is ruled out 
by using method 1, but the use of method 2, which is much
quicker to implement, can vastly reduce the region of parameter space
that one must explore with method 1. There may also be theories where
method 1 or its variations\footnote { We have suggested a different
method of reduction in a related context in ref. \cite {dg}.}
can provide superior bounds than method 2. 
A simple variation of method 2 is obtained 
by replacing the straight line in $R^m$ by a
curve. When several intuitive choices for the reduction are available, 
it is difficult to say which method will provide superior results.
However method 2 is always the quickest one and should be used to 
rule out as much of the parameter space as possible. 

\section {$d=1,2$}

We would like to remark on the cases $d=1$ and $d=2$. Statement 2
is in fact generically true for $d=2$. To see this consider the
deformation of the functional $T$ in $S$ to $T(\eta)$ defined by
\be 
T(\eta) = {1 \over 2} \int d^2x\left ({\partial \phi_a \over \partial
  x_i}\right )^{2-\eta},
\label {deform}
\ee
where $\eta$ is a positive real number. With this deformation the
equation (\ref {tv}) is modified for ($d=2$) to 
\be
\eta \, T(\eta) + 2 \, V =0.
\ee
Statement 2 is true as $\eta$ approaches zero from the positive
direction. Thus $G$ remains a saddle point in the sense that it is a
maximum of the action with respect to  some variation orthogonal to $C_1$. However
the negative eigenvalue of the Hessian ${\partial ^2 S \over \partial
\phi_a^2}$ may approach zero as 
$\eta \to 0$. But there is 
no underlying symmetry to make this situation 
generic. In other words, the eigenvalues of the Hessian are
likely to be of the order of $m^2$ where $m$ is some
mass scale in the theory. 
Unless there is a symmetry to protect its smallness, an 
eigenvalue can not be made zero without fine tuning the parameters of
the theory. 

The case $d=1$ can be treated as follows. The equations of motion to be
satisfied by the bounce are
\be
{\partial ^2 \phi_a \over \partial t^2}= {\partial U \over \partial \phi_a}
\label {done}
\ee
with the boundary conditions $\phi_a (\pm \infty) \to 0\,$, ${\partial 
  \phi_a \over \partial t}|_{t=0}=0$. Solutions to (\ref {done}) resemble
  the motion of a particle of unit mass in the potential
  $-U(\phi_a)$. Let us 
define the surface $D$ in the $m$ dimensional space of
  the ``fields'' $\phi_a$ as the surface where $U=0$. The space $D$
  seperates as $D=D_0 \oplus D_1$ with $D_0$ given by the part that is
  connected to the point
  $\phi_a=0$. By definition 
 $D_0$ and $D_1$ are not connected in $D$. Consider
  the space of $Z_2$ invariant trajectories $\phi_a(t) = \phi_a(-t)$
with  $\phi_a(0) \in D_1$,  $\phi_a(\infty) \to D_0$ and
satisfying the constraint $U[\phi_a(t)]\ge 0$. The action of these
trajectories is positive definite and has a global minimum, which, we
claim, is generically a bounce. 

Let us briefly substantiate the claim which can also be proven along the
lines of ref. \cite {coleman2}. $G$ is a minimum with respect to  all
allowed variations except variations that move the point $\phi_a(0)$ out
of the surface $D_1$. Generically, connected parts in $D_1$ are $m-1$
dimensional surfaces. This leaves a single variation that moves
$\phi_a(0)$ out of $D_1$. We need to show that there is a direction of
instability associated with this variation.
The entire trajectory is a solution of the
equations of motion. By conservation of the ``energy'' $E=\sum _{a=1}^m
\left [{1 \over 2} ({d\phi_a
  \over dt})^2\right ] - U(\phi_1, \phi_2 ... \phi_m)$, 
the total energy is zero on this
trajectory and the velocity is zero at the ``highest point''
($ {d\phi_a \over \ dt}(0)=0$). This implies that the
generator of time translations ${d\phi_a \over dt}$ has a node at
$t=0$. By time translational invariance of the bounce, 
the eigenvalue of the Hessian
$ {\partial ^2 S\over \partial \phi_a^2}$ that corresponds to time translations
is zero. Therefore there is a nodeless variation with a negative
eigenvalue and $G$ is a bounce.

{{\bf {Acknowledgements}}}: 
I am grateful to Claudio Rebbi for many useful discussions.
I also thank Bogdan Dobrescu for some key ideas, 
Andrew Cohen, Alexander Kusenko, Martin Schmaltz, Andreas Wipf 
and Yuri Sarid for valuable
suggestions and Bogdan Dobrescu and Lisa Randall for the permission to
use some data from ref. \cite {ddr}. 
This work was supported by the Department of Energy under the grant
DE-FG02-91ER40676.

\vfill

\end{document}